\documentclass[fleqn,10pt]{wlscirep}
\title{Spin filtering effect generated by the inter-subband spin-orbit coupling in the bilayer nanowire
with the quantum point contact}

\author[1,*]{Pawe{\l} W\'ojcik}
\author[1]{Janusz Adamowski}
\affil[1]{AGH University of Science and Technology, Faculty of
Physics and Applied Computer Science, al. Mickiewicza 30,
Krak\'ow, Poland}

\affil[*]{pawel.wojcik@fis.agh.edu.pl}


\begin{abstract}
The spin filtering effect  in the bilayer nanowire with quantum point contact is 
investigated theoretically. We demonstrate the new mechanism of the spin filtering based on the
lateral inter-subband spin-orbit coupling, which for the bilayer nanowires has been reported
to be strong. The proposed spin filtering effect is explained as the joint effect of the
Landau-Zener inter-subband transitions caused by the hybridization of states with opposite spin
(due to the lateral Rashba SO interaction) and the confinement of carriers in the quantum point
contact region.
\end{abstract}
\begin{document}

\flushbottom
\maketitle

\thispagestyle{empty}

\section*{Introduction}
A fabrication of a controllable source of a spin polarized current that operates without a magnetic
field is one of the most important challenges of semiconductor spintronics\cite{Fabian2007}. Among
spin filters proposed over the years, including these based on carbon nanotubes\cite{Hauptmann2008},
quantum dots\cite{Folk2003}, Y-shaped nanostructures\cite{Wojcik2015,Tsai2013} or resonant tunneling
diodes\cite{Wojcik2012,Wojcik2013,Slobodsky2003}, recently, the special attention is paid to the
quantum point contacts (QPC) with the spin-orbit (SO)
interaction\cite{Debray2009,Das2011,Das2012,Bhandari2012,Bhandari2013,Wan2009,Kohda2012,Nowak2013,Ngo2010}. 
In such nanostructures, the spin filtering effect results from the
interplay between the SO coupling\cite{Rashba1984,Desselhaus1955} and the quantum confinement.
Recent papers\cite{Debray2009,Das2011,Das2012,Bhandari2012,Bhandari2013,Wan2009,
Kohda2012} have reported the experimental evidence of the spin filtering in QPCs, which manifests
itself as the plateau of conductance at $0.5 G_0$ $(G_0=2 e^2 /h)$ measured in the absence of the
external magnetic field. The $0.5G_0$ plateau has been explained as resulting from the combination
of the three effects\cite{Ngo2010,Bhandari2013,Wan2009}: an asymmetric lateral confinement, a
lateral Rashba SO interaction, and an electron - electron interaction. More precisely, the asymmetry
in a lateral confinement, induced by the different voltages applied to the side electrodes of QPC,
is a source of a lateral electric field. Due to the SO interaction, this electric field, in the
electron's rest frame, is seen as an effective magnetic field, which initializes an imbalance
between the spin-up and spin-down electrons. As shown by Ngo et al.\cite{Ngo2010} 
this so-called lateral Rashba SO interaction  leads to the low spin
polarization of the current not exceeding 6 \%, and therefore its presence did not explain the
$0.5G_0$ plateau observed in the experiments. The full explanation has been given by the further
studies, which have shown that the predicted weak spin filter effect\cite{Ngo2010} can be
strengthened by the electron-electron interaction leading to the nearly 100~\% spin polarization in
the regime of the single-mode transport\cite{Bhandari2013}. QPCs with the SO interaction have been
successfully used as the spin injector and detector in the recent experimental realization of the
spin transistor\cite{Chuang2015, Alomar2016}, in which about $10^5$ times greater conductance
oscillations have been observed as compared to the conventional spin-field effect transistor based
on ferromagnets\cite{Koo2009,Wojcik2014}. 

The experimental realizations of spin filters based on QPCs, reported so far, are based on a
two-dimensional electron gas (2DEG) confined in the narrow quantum well at the
AlGaAs/GaAs or InAs/InAlAs interface\cite{Das2011,Das2012}, in which the
electrons occupy the lowest-energy state ("single occupancy"). However, recently, wider or coupled
quantum wells with two populated subbands ("double occupancy") have attracted a growing interest
of both experimentalists\cite{Fisher2005,Fisher2006,Bentmann2012,Hernandez2013,Hu1999,Akabori2012}
and
theoreticians\cite{Calsaverini2008,Bernardes2007,Fu2015,Chwiej2016PE,Chwiej2016PB,
Chwiej2016PRB}. In this case, we deal with the vertically coupled nanowires, in which the coupling
strength is determined by the wave function overlap  between the ground and first excited states.
The additional orbital degree of freedom in the bilayer nanowires leads to
interesting physical effects such as inter-subband induced band
anticrossing and spin mixing\cite{Bentmann2012}. 
The SO interaction in quantum well with double occupancy has been studied by Bernardes et al\cite{Bernardes2007}. 
The inter-subband induced SO interaction has been found which
results from the coupling between states with opposite parity. It can give raise to intriguing 
physical phenomena, e.g. unusual
Zitterbewegung\cite{Bernardes2007} or intrinsic spin Hall effect in symmetric quantum
well\cite{Hernandez2013}. The influence of the inter-subband SO interaction in
bilayer nanowire on the spin transistor action has been studied in our
recent paper\cite{Wojcik2016_arxiv}. We have shown that the resonant behavior of spin-orbit coupling
constants obtained for zero gate voltage leads to the spin transistor operation, in which
the on/off transition should be realized in the narrow voltage range.

In the present paper, we demonstrate the novel mechanism of spin filtering based on the
lateral inter-subband spin-orbit coupling in the bilayer nanowire with QPC. We find
that for the non-zero inter-subband coupling induced by the  lateral Rashba SO interaction, the spin
polarization of the current flowing through the QPC is almost full. We analyze the conditions, under
which this polarization takes place. The observed spin filtering effect is explained as the joint
effect of the Landau-Zener inter-subband transitions caused by the hybridization of states with
opposite spins and the quantum confinement in the QPC region. Our results provide a new
mechanism to implement spin-polarized electron sources in the realistic bilayer nanowires which
can be built from the double quantum well or wide quantum well structure.

\section*{Theoretical model}
\subsection*{Model of the nanostructure}
\label{sec2}
\begin{figure}[ht]
\begin{center}
\includegraphics[scale=0.5, angle=0]{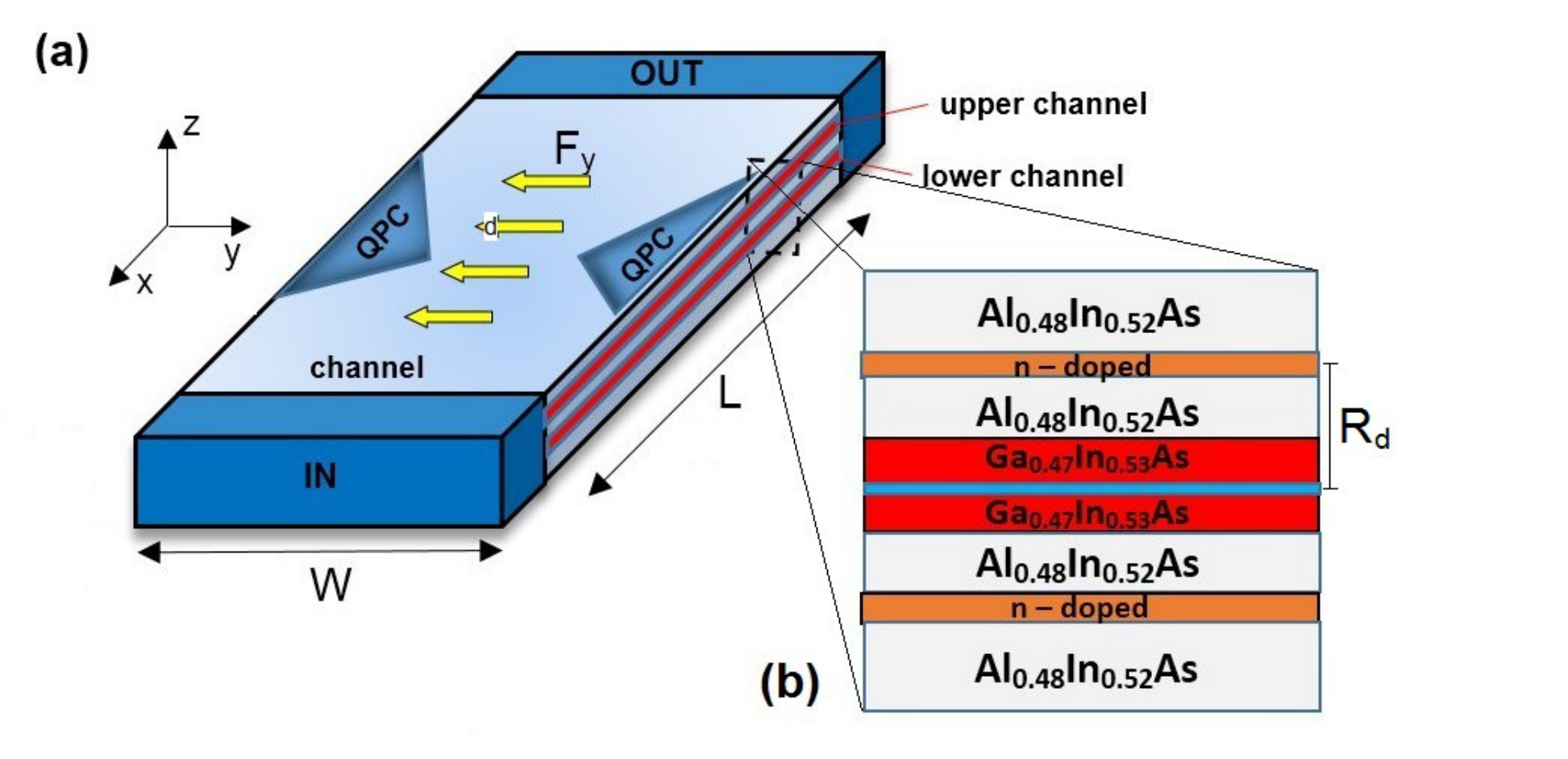}
\caption{
(a) Schematic of the bilayer nanowire with the QPC. In the presence of the lateral electric field
$F_y$, the unpolarized current injected from the contact IN, after passing through
the QPC, is almost fully spin polarized. (b) Cross section of the exemplary realization of the
bilayer nanowire based on the Al$_{0.48}$In$_{0.52}$As/Ga$_{0.47}$In$_{0.53}$As double quantum
wells.}
\label{fig1}
\end{center}
\end{figure}

We consider the bilayer nanowire consisting of two coupled conducting channels of width $W$ and
length $L$ (Fig.~\ref{fig1}). Both ends of the nanowire are connected to the reflectionless, ideal
leads denoted as IN and OUT. In the middle of the nanowire, the QPC is located as schematically
presented
in Fig.~\ref{fig1}(a). 
In recent experiments\cite{Fisher2005,
Fisher2006,Bentmann2012,Hernandez2013,Hu1999,Akabori2012}, vertically stacked and coupled nanowires
with QPC's  are prepared from 2DEG bilayer systems built from double quantum wells or a wide
quantum well in which the weak Coulomb repulsion gives rise to the "soft" barrier in the middle of
the quantum well.  
Figure~\ref{fig1}(b) presents the cross-section of the exemplary double quantum well
heterostructure, which consists of two
Al$_{0.48}$In$_{0.52}$As/Ga$_{0.47}$In$_{0.53}$As quantum wells with a central
Al$_{0.3}$In$_{0.7}$As barrier with width $w_b$, which controls the coupling between the
conduction electron states in the quantum wells. For the sufficiently high central barrier,
the quantum wells are separated and the electron wave functions are localized in one of
the quantum wells. For the symmetric structure each of the states is fourfold degenerate whereby the
twofold degeneracy results from the spin states and twofold degeneracy is related to the
geometric symmetry. If the height of the barrier decreases, the states in the quantum wells become
coupled to each other with the coupling strength determined by the wave function overlap. The
formation of the symmetric and antisymmetric (bounding and anti-bounding) states leads to the
splitting of the previously degenerate
quantum states with the splitting energy in the range $1-10$~meV\cite{Fisher2005,Fisher2006}. For
appropriate electron density only the two spin degenerate subbands (the ground and first excited)
are occupied (double occupancy) leaving the rest of the higher-energy subbands unoccupied since
their energy is considerably higher than the energy of the ground and first excited states. 

The spin filtering effect proposed in this paper requires the lateral Rashba
SO interaction i.e. SOI generated by the lateral electric field
$\mathbf{F}=(0,F_y,0)$. In fact, recently, Gvozdic and Ekenberg\cite{Gvozdic2007}  pointed out that
in the modulation-doped wide or coupled quantum wells, used for the bilayer nanowires fabrication,
the large intrinsic $F_y$ exists. Alternatively, in the experiment, the lateral electric field
can also be generated by the side gates attached to the channel. Regardless of the origin, the
lateral electric field $F_y$ induces the Rashba SO interaction with the effective magnetic field
directed along the grown $z$-axis.  The possibility of using the SO interaction
induced by the lateral electric field has been recently reported in many
experiments\cite{Chuang2015,Debray2009,Das2011}, in which the lateral SO coupling constant has
been reported to vary in the range $0.04$~eV\AA$-0.5$~eV\AA.\\

\subsection*{Numerical methods}
Let us define the four element basis $\{ |1,\uparrow\rangle , |1,\downarrow\rangle , 
|2,\uparrow\rangle , |2,\downarrow\rangle\}$ which consists of
the spin-degenerate ground and first excited eigenstate related to the
confinement in the $z$ direction. In the presence of the SO interaction, we can derive the $4
\times 4$ Hamiltonian in the basis of these states\cite{Calsaverini2008,Wojcik2016_arxiv}. 
After introducing a set $\pmb{\tau}$ of Pauli-like matrices in the orbital space, the Hamiltonian of
the system takes on the form (full derivation of Hamiltonian (\ref{eq:H}) can be found in
Supplementary)
\begin{eqnarray}
\label{eq:H}
H&=&\left [ \frac{\hbar ^2 (\hat{k} _x^2+\hat{k} _y^2)}{2m^*} + U_{QPC}(x,y) + |e|F_yy +
\varepsilon _+ \right ] \mathbf{1} \otimes \mathbf{1}   
- \varepsilon _{-}\tau _z \otimes \mathbf{1} +\beta |e| F_y \hat{k}_x \mathbf{1} \otimes \sigma _z 
- \beta \delta |e| F_y \tau _y \otimes \sigma _x \nonumber \\
 &+&\beta_{12} \tau_{x} \otimes \left ( \sigma _x \hat{k}_y - \sigma _y \hat{k}_x \right ),
 \end{eqnarray}
where $\mathbf{1}$ is the $2 \times 2$ unit matrix, $m^*$ is the electron effective mass, $\hat{k}
_{x(y)}=-i \partial / \partial x(y)$ is the wave vector operator, $\beta$ is
the lateral Rashba spin-orbit coupling constant, $\beta_{12}$ is the inter-subband lateral
Rashba spin-orbit coupling constant, $\delta=\langle 1, \sigma | \partial / \partial z | 2, \sigma
\rangle $ is the inter-subband coupling constant, $\varepsilon _{\pm}=(\varepsilon _2 \pm
\varepsilon _1)/2$ with $\varepsilon _{1(2)}$ being
the energy of eigenstate $|1, \sigma \rangle$ ($|2, \sigma \rangle$), $U_{QPC}(x,y)$ is the
potential energy of electron from the QPC which is given by
\begin{eqnarray} 
&&U_{QPC}(x,y)=V_{QPC} \exp \left [ - \left ( \frac{x-L/2 }{\xi_x/2} \right ) ^2 \right ]  \times
\left \{ \exp \left [ - \left ( \frac{y-W/2 }{\xi _y/2} \right ) ^2 \right ] +  \exp \left [
- \left ( \frac{y+W/2 }{\xi _y/2} \right ) ^2 \right ] \right \}, 
\end{eqnarray}
where $V_{QPC}$ is the maximal potential energy in the QPC while $\xi _x$ and $\xi _y$ determine the
extension of the QPC in the $x$ and $y$ directions, respectively.
In Hamiltonian (\ref{eq:H}) we neglect the intra-subband SO coupling assuming that the system is
symmetric in the $z$ direction with respect to the reflection $z\rightarrow
-z$\cite{Wojcik2016_arxiv}. Since the Dresselhaus SO coupling constant $\gamma _D \sim 1/d_{QW}^2$ ($d_{QW}$
is the width of the quantum well in the $z$ direction), for the wide quantum well used in the experimental 
realization of  bilayer nanowires, the strength of the Dresselhaus SO interaction is a few orders of magnitude smaller than the Rashba SO coupling. 
This allows us to neglect the Dresselhaus term in the Hamiltonian (\ref{eq:H}) - detailed discussion of the Dresselhaus SO coupling 
in the considered nanostructure and its influence on the presented spin filtering can be found in Supplementary material. 

The calculations of the  conductance have been performed by the scattering matrix
method using the Kwant package\cite{kwant}. For this purpose we have transformed the
Hamiltonian (\ref{eq:H}) into the discretized form on the grid $(x_{\mu}, y_{\nu})= (\mu dx, \nu 
dx$) with $\mu, \nu = 1,2, \ldots$, where $dx$ is the lattice constant.
We introduce the discrete representation of the electron state in the $4 \times 4$ space as follows:
$|\Psi(x_{\mu}, y_{\nu})\rangle
=
\left(|\psi_1^{\uparrow}( x_{\mu},y_{\nu})\rangle
,|\psi_1^{\downarrow}( x_{\mu},y_{\nu})\rangle, |\psi_2^{\uparrow}( x_{\mu},y_{\nu})\rangle,
|\psi_2^{\downarrow}( x_{\mu},y_{\nu})\rangle \right)^T
= |\Psi_{\mu, \nu}\rangle$.
The Hamiltonian (\ref{eq:H}) takes on the discretized two-dimensional form 
\begin{eqnarray}
\label{HTB}
\mathcal{H}_{2D}&=& \sum\limits_{\mu\nu}  \big [ (4t + \varepsilon _+ + U_{QPC}(x_{\mu},y_{\nu}) +
|e|F_yy_{\nu}) \mathbf{1} \otimes \mathbf{1} 
- \varepsilon _- \tau _z \otimes \mathbf{1} - \beta \delta |e| F_y \tau _y \otimes \sigma _x \big ] |
\Psi_{\mu,\nu } \rangle \langle \Psi_{\mu,\nu}|  \\
 &+& \sum _{{\mu}{\nu}} \left [ t \mathbf{1} \otimes \mathbf{1} + it_{SO} \beta_{12} \tau
_x \otimes \sigma _y - it_{SO} \beta |e| F_y \mathbf{1} \otimes \sigma _z\right ] | \Psi_{\mu+1,\nu }
\rangle \langle \Psi_{\mu,\nu}| + H.c. \nonumber
\\
 &+& \sum _{{\mu}{\nu}} \left [ t \mathbf{1} \otimes \mathbf{1} - it_{SO} \beta_{12} \tau
_x \otimes \sigma _x \right ] | \Psi_{\mu,\nu+1 } \rangle \langle \Psi_{\mu,\nu}| +H.c., \nonumber
\end{eqnarray}
where $t=\hbar ^2/(2m dx^2)$ and $t_{SO}=1/(2dx)$. \\
In the calculations we assume the hard-wall boundary 
conditions in the $y$ direction i.e. $ |\Psi(x,0)\rangle=|\Psi(x,W)\rangle=0 $. In the $x$-direction,
the boundary conditions are set based on the fact that the electron state in the input is a linear combination of 
the state in which the electron is injected into the channel and all possible states in which the electron 
can be reflected from the QPC. Therefore
\begin{equation}
 |\Psi_{in}(0,y)\rangle = c_{k_{n}} ^{\sigma} \exp (i k_{n}^{\sigma} x) | \varphi_{in,n}^{\sigma} (y) \rangle + \sum _{m,\sigma'}
c_{k_m}^{\sigma'} \exp (-ik_m^{\sigma'} x) |\varphi_{in,m}^{\sigma'}(y)\rangle,
\end{equation}
where $n,m=1,2$, $\sigma,\sigma '=\uparrow, \downarrow$ and $| \varphi_{in,n}^{\sigma} (y) \rangle$ are the eigenstates of the hamiltonian (\ref{HTB})
calculated 
by the use of the boundary conditions in the input ($\mu=0$)
\begin{equation}
 |\Psi_{\mu \pm 1, \nu } \rangle = \exp (\pm i k dx) |\Psi_{\nu} \rangle.
\end{equation}
Accordingly, the boundary conditions for the output are given by
\begin{equation}
 | \Psi _{\mu, \nu+1} \rangle = \exp (ikdx) | \Psi _{\mu, \nu} \rangle.
\end{equation}

Let us assume that the electron with spin $\sigma$ in the subband $n$ ($n=1,2$)  is injected from
the input contact into the nanowire. The electron can be transmitted through the QPC
in one of the four possible processes (in parenthesis the symbol of the transmission probability
is given): intra-subband transmission with spin conservation $(T_{nn}^{\sigma \sigma})$,
intra-subband transmission with spin-flip $(T_{nn}^{\sigma \bar{\sigma}})$,  inter-subband
transmission with spin conservation ($T_{nm}^{\sigma \sigma}$), and inter-subband transmission with
spin flip ($T_{nm}^{\sigma \bar{\sigma}}$), where $\bar{\sigma}$ denotes the spin opposite to
$\sigma$, while $T_{nm}^{\sigma \sigma'}$ is the probability of the electron transmission between
the subbands $|n,\sigma \rangle \rightarrow |m,\sigma' \rangle $, ($n,m=1,2$, $\sigma,\sigma
'=\uparrow, \downarrow$).

Having determined the transmission coefficients $T^{\sigma \sigma '}_{nm}$
we calculate the conductance in the ballistic regime using the Landauer formula
\begin{equation}
 G_{nm}^{\sigma \sigma^{\prime}}=\frac{e^2}{h} \int_0^{\infty} T_{nm} ^{\sigma \sigma ^{\prime}} (E) \left (
\frac{\partial f_{FD}(E,E_F)}{\partial E} \right ) dE,
\end{equation}
where $f_{FD}(E,E_F)=1/[1+\exp(E-E_F)/k_BT]$ is the Fermi-Dirac distribution function, $T$ is
the temperature and $E_F$ is the Fermi energy.
The spin-dependent conductances through the nanostructure are given by 
\begin{eqnarray}
G^{\uparrow}&=&\sum _{n,m=1}^{2} (G_{nm}^{\uparrow \uparrow} + G_{nm}^{\downarrow \uparrow}), \\
G^{\downarrow}&=&\sum _{n,m=1}^{2} (G_{nm}^{\uparrow \downarrow} + G_{nm}^{\downarrow \downarrow}).
\end{eqnarray}
The total conductance $G=G^{\uparrow}+G^{\downarrow}$ and the spin polarization of current
\begin{equation}
P=(G^{\uparrow} - G^{\downarrow})/(G^{\uparrow}+G^{\downarrow}). 
\end{equation}
The conductance calculations have been performed for  $dx=2$~nm,
$L=3000$~nm, $W=92$~nm, $\xi _x=300$~nm, $\xi _y=48$~nm and
$V_{QPC}=12$~meV. We use the material parameters corresponding to In$_{0.5}$Ga$_{0.5}$As, 
i.e., $m_e=0.0465m_0$ and the Rashba spin-orbit interaction constant $|e|\beta F_y=10$~meVnm\cite{Das2011}.
The energy difference between the two occupied subbands is taken to be $\Delta \varepsilon =
\varepsilon _2 - \varepsilon_1=1$~meV.

\section*{Results and Discussion}
In this section, we present the results of the conductance calculations and explain the
physical mechanism responsible for the nearly full spin polarization obtained in the
bilayer nanowires with QPC.
First, we assume that the inter-subband induced SO coupling constant $\beta_{12}=0$. 
\begin{figure}[ht]
\begin{center}
\vspace{0.5cm}
\includegraphics[scale=0.6, angle=0]{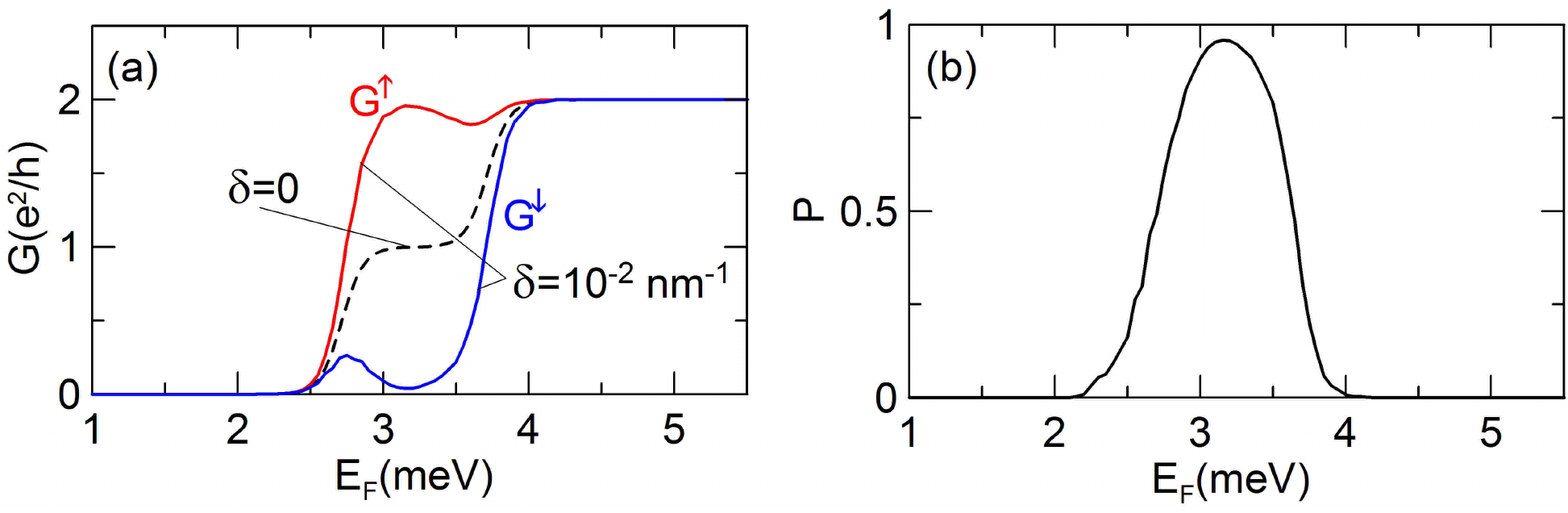}
\caption{
(a) Conductance $G$ as a function of Fermi energy $E_F$ for two values of inter-subband coupling
constants $\delta$. Red and blue curves correspond to $G^{\uparrow}$ and
$G^{\downarrow}$, respectively. Black dashed curve shows the results for $\delta=0$, for which
$G^{\uparrow}=G^{\downarrow}$. (b) Spin polarization $P$ of the current as a function of Fermi
energy $E_F$ for $\delta=10^{-2}$~nm$^{-1}$. }
\label{fig2}
\end{center}
\end{figure}
\begin{figure}[ht]
\begin{center}
\includegraphics[scale=0.6, angle=0]{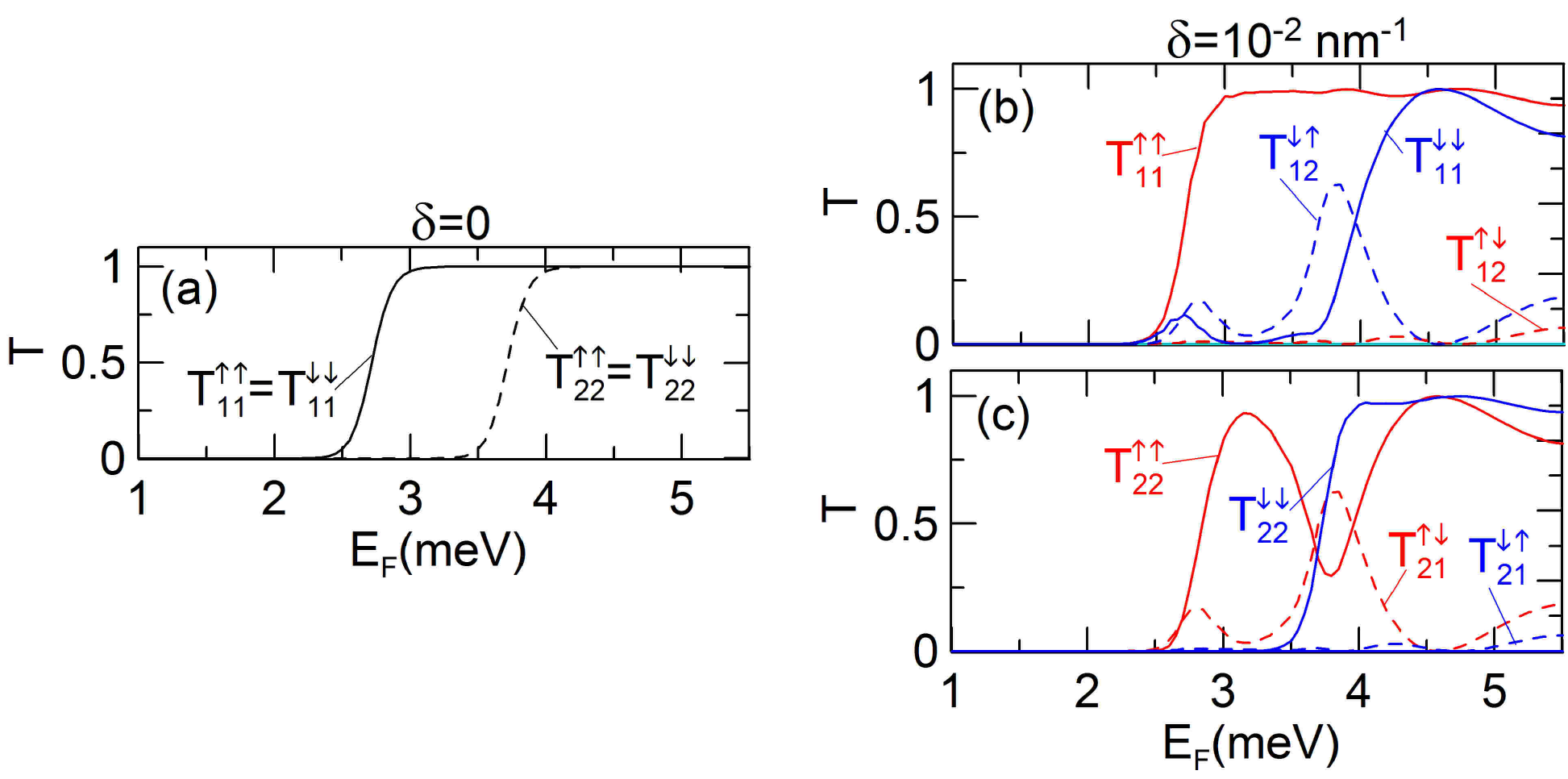}
\caption{
Transmission probabilities $T_{nm}^{\sigma \sigma'}$ versus Fermi energy $E_F$ for (a)~$\delta=0$
and (b,c)~$\delta=10^{-2}$~nm$^{-2}$ for the electron injected into (b) the first
subband $|1,\sigma \rangle$ and (c) the second subband $|2,\sigma \rangle$,
$\sigma=\uparrow,\downarrow$.}
\label{fig3}
\end{center}
\end{figure}
Fig.~\ref{fig2}(a) displays the conductance $G^{\uparrow \downarrow}$ as a function of the Fermi
energy $E_F$ for two values of the inter-subband coupling constant $\delta$. Red and blue curves
correspond to $G^{\uparrow}$ and $G^{\downarrow}$, respectively. For $\delta=0$, depicted by the
black dashed curve, $G^{\uparrow}=G^{\downarrow}$ hence the spin polarization of current $P=0$ in
entire range of the Fermi energy. The two conductance steps are due to the subsequent
subbands passing through the Fermi level in the QPC region, each contributing to the increase of the
conductance by $e^2/h$. 
For the nonzero inter-subband coupling, i.e for $\delta=10^{-2}$~nm$^{-1}$, the conductances
$G^{\uparrow}$ and $G^{\downarrow}$  differ from each other in some range of the Fermi energy, which
leads to the almost full spin polarization of current presented in Fig.~\ref{fig1}(b). 
We note that the spin polarization occurs only in the Fermi energy range, which
corresponds to the conductance step for $\delta=0$. The transmission probabilities
$T_{nm}^{\sigma \sigma'}$ for $\delta=0$ and $\delta=10^{-2}$~nm$^{-1}$ are presented in
Fig.~\ref{fig3}.
Comparing results in Fig.~\ref{fig3}(a) and Fig.~\ref{fig3}(b,c) we see that the imbalance
between the spin-up and spin-down conductance, in the Fermi energy range, in which the spin
polarization is observed, results from the suppression of the intra-subband transmission
of spin-down electrons in the first subband $T_{11}^{\downarrow \downarrow}$ and the enhancement of
the intra-subband transmission of spin-up electrons $T_{22}^{\uparrow \uparrow}$. As expected the
inter-subband transmissions reveal the symmetry $T_{12}^{\uparrow \downarrow}=T_{21}^{\downarrow
\uparrow}$ and $T_{12}^{\downarrow \uparrow}=T_{21}^{\uparrow \downarrow}$.

\begin{figure}[t]
\begin{center}
\includegraphics[scale=0.6, angle=0]{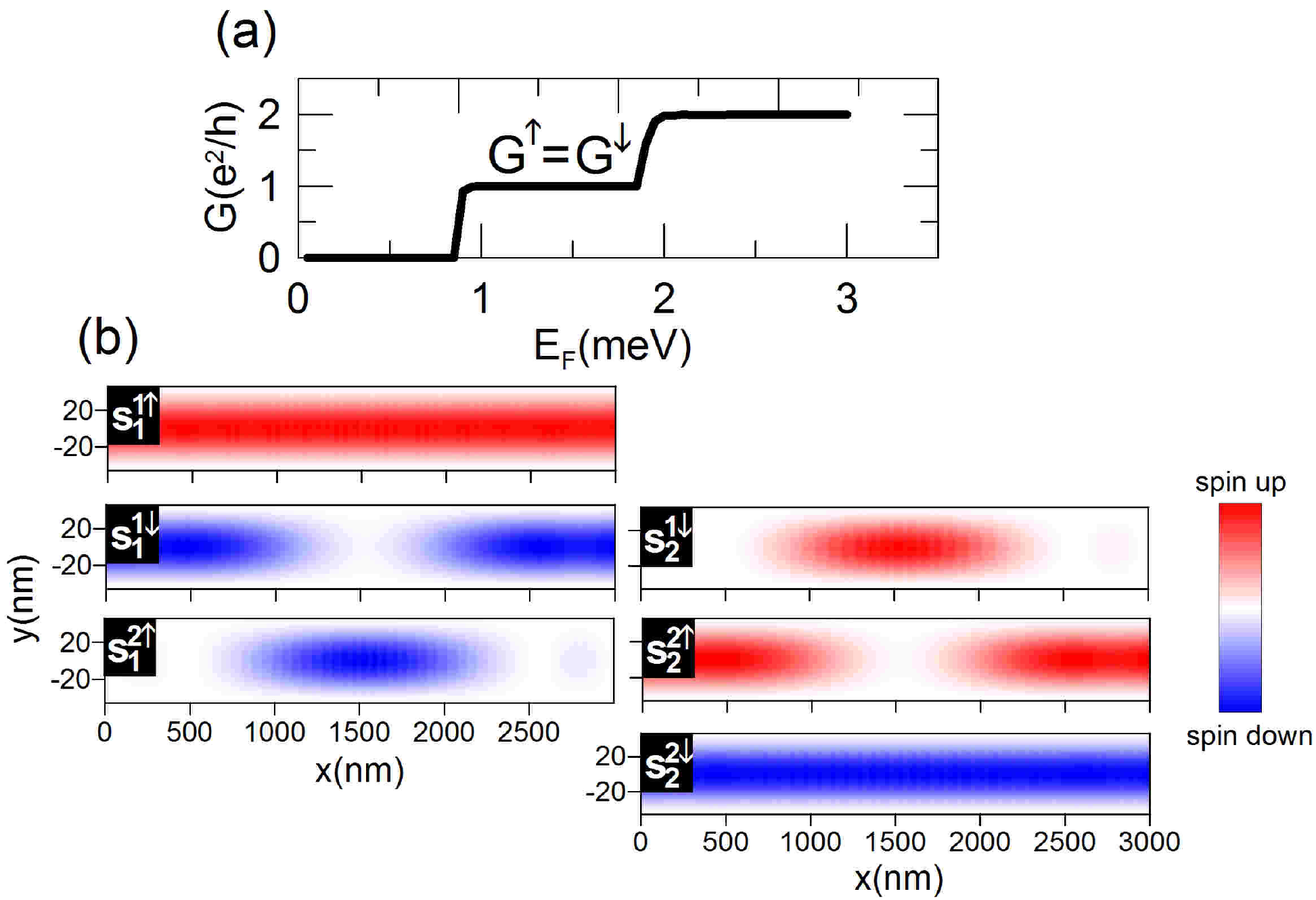}
\caption{
(a) Conductance $G$ as a function of Fermi energy $E_F$ for $V_{QPC}=0$. (b) The $z$ component of
the partial spin density distributions $s_{m}^{n\sigma}$, where $m$ is the index of the subband, for
which the spin density distribution is presented, while $n\sigma$ denotes the index of the subband,
including spin, from which the electrons are injected into the nanowire. Results for $E_F=3.15$~meV
and $V_{QPC}=0$. }
\label{fig4}
\end{center}
\end{figure}

\begin{figure}[h!]
\begin{center}
\includegraphics[scale=0.6, angle=0]{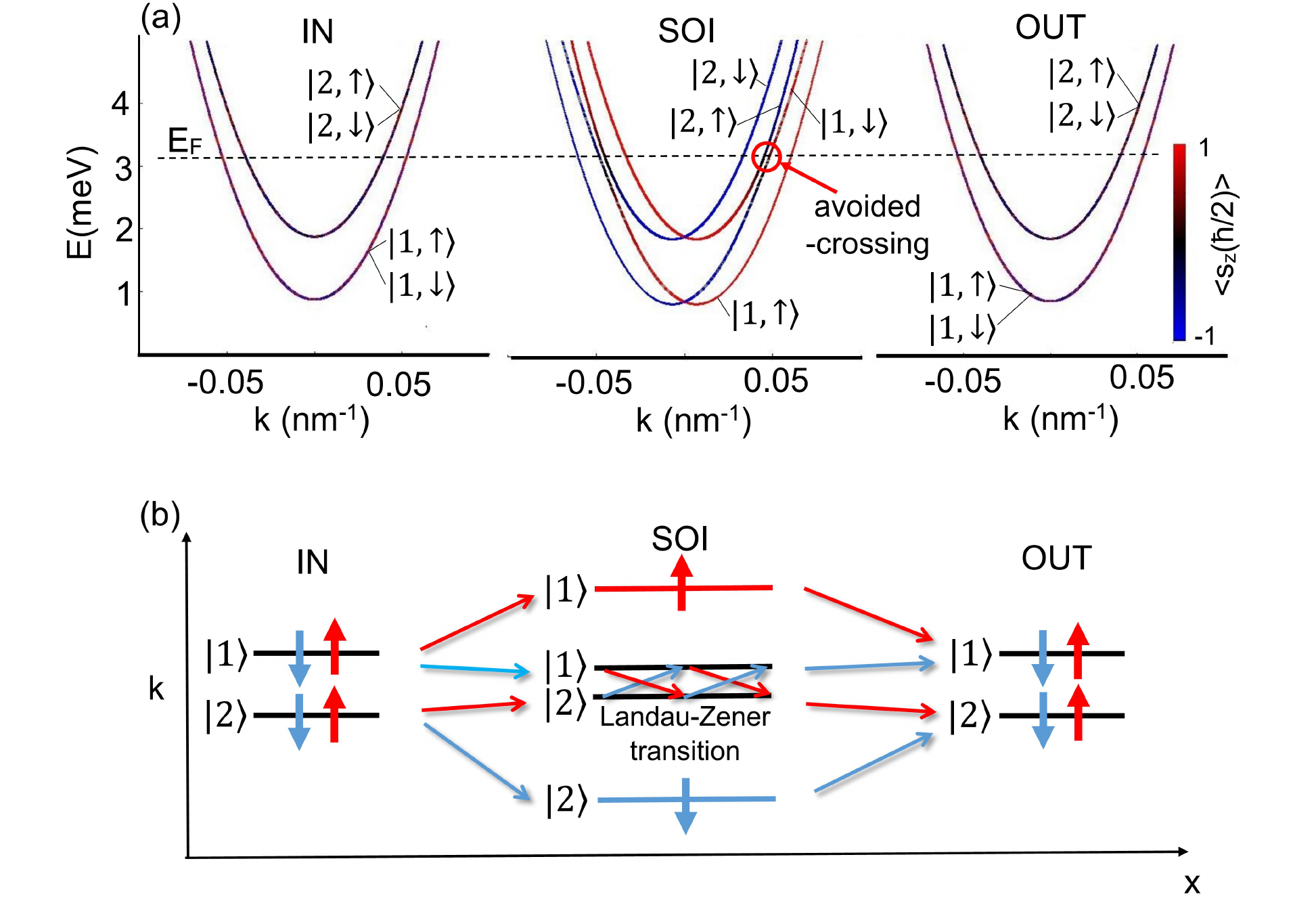}
\caption{ (a) Dispersion relations $E$ vs $k$ in the leads IN and OUT as well as in
the nanowire with the SO interaction. The dashed horizontal line denotes the Fermi
energy $E_F=3.15$~meV, for which the spin density distribution maps are presented in
Fig.~\ref{fig4}(b). (b) Schematic illustration of the possible transmission processes through the
nanowire without QPC.
 }
\label{fig5}
\end{center}
\end{figure}

In order to explain the physical mechanism behind the spin filtering effect let us first
consider the nanowire without QPC ($V_{QPC}=0$).
As shown in Fig.~\ref{fig4}(a), which presents the spin-dependent conductance
$G^{\uparrow \downarrow}(E_F)$, the constriction in the form of QPC located in the nanowire is
necessary to obtain the spin filtering. Without QPC, $G_{\uparrow}=G_{\downarrow}$, which leads 
to the spin polarization $P=0$. The full understanding of the spin filtering mechanism, which
emerges when we add QPC, requires the understanding of the spin dynamics in the
nanowire without the constriction.
Fig.~\ref{fig4}(b) presents the $z$ component of the partial spin density distributions
$s_{m}^{n\sigma}$, where $m$ is the index of the subband, for which the spin density distribution is
presented, while $n\sigma$ denotes the index of the subband, including spin, from which the
electrons are injected into the nanowire. Explicitly, 
\begin{eqnarray}
 s_{1}^{n\sigma}(x_{\mu},y_{\nu})&=&|\psi_1^{\uparrow}( x_{\mu},y_{\nu})|^2-|\psi_1^{\downarrow}(
x_{\mu},y_{\nu})|^2, \\
 s_{2}^{n\sigma}(x_{\mu},y_{\nu})&=&|\psi_2^{\uparrow}( x_{\mu},y_{\nu})|^2-|\psi_2^{\downarrow}(
x_{\mu},y_{\nu})|^2.
\end{eqnarray}
The dispersion relations for the leads IN and OUT as well as in the central part of the nanowire
with the SO interaction are presented in Fig.~\ref{fig5}(a). The Fermi energy, for which the maps
from Fig.~\ref{fig4}(b) have been calculated, is marked by the dashed horizontal line. We set
$E_F=3.15$~meV which corresponds to the maximal spin polarization of current presented in
Fig.~\ref{fig2}(b).

As shown in Fig.~\ref{fig4}(b) the electrons injected in the states $|1,\uparrow \rangle$ and
$|2, \downarrow \rangle$ flow through the nanowire in the subband into which they were injected,
conserving their spin. Different behavior is observed for the electrons injected in the
states $|1,\downarrow \rangle$ and $|2, \uparrow \rangle$. The electrons
injected into the subband $|1,\downarrow \rangle$ ($|2,\uparrow \rangle$) are transmitted to the
state $|2, \uparrow \rangle$ ($|1,\downarrow \rangle$) in the middle of the nanowire and are back in
their original state before leaving the nanowire through the lead OUT. To explain this
behavior let us consider the electron propagating in the state $|1,\downarrow \rangle$ from the
input channel IN. In the nanowire where the lateral Rashba SO interaction is present, the spin is no
longer a good quantum number, since the orbital and spin degrees of freedom are mixed. For the
appropriate Fermi energy the states $|1,\downarrow \rangle$ and $|2, \uparrow \rangle$ hybridize
giving raise to the
avoided crossing in the spin-split subbands presented in Fig.~\ref{fig5}(a). The probability
of the transition through the avoided crossing depends on the degree of the adiabaticity of the
electron transport and the avoided crossing width as predicted by the
Landau-Zener theory\cite{Landau,Zener}. For the fully diabatic transport the probability of
electron transfer from the state $|1,\downarrow \rangle$ to $|2, \uparrow \rangle$ is equal to $1$.
Therefore, the transfer probability through the avoiding crossing depends on the rate of the energy
level changes when the electron from the contact, in which the SO interaction is absent, flows
through the nanowires with the SO interaction. In our case, this effect is implemented by the
spatially dependent SO coupling constants which vary in accordance with the function 
\begin{equation}
f(x)=\exp {\left [ - \left ( \frac{x-L/2}{\xi _{SO} /2 } \right )^{2p} \right ]},
\end{equation}
where $\xi_{SO}\approx L$ and $p$ is the so-called "softness" parameter. We assume $p=10$, which
guarantees the diabatic transport between the leads and the nanowire. \\ 
The schematic illustration of all possible electron transmission processes through the nanowire in
the $k$-space is presented in Fig.~\ref{fig5}(b). Among them the Landau-Zener transition between the
states $|1,\downarrow \rangle$ and $|2, \uparrow \rangle$ (with spin flip) is a key to understanding
the spin filtering effect, which emerges when we introduce the QPC. 

\begin{figure}[h!]
\begin{center}
\includegraphics[scale=0.6, angle=0]{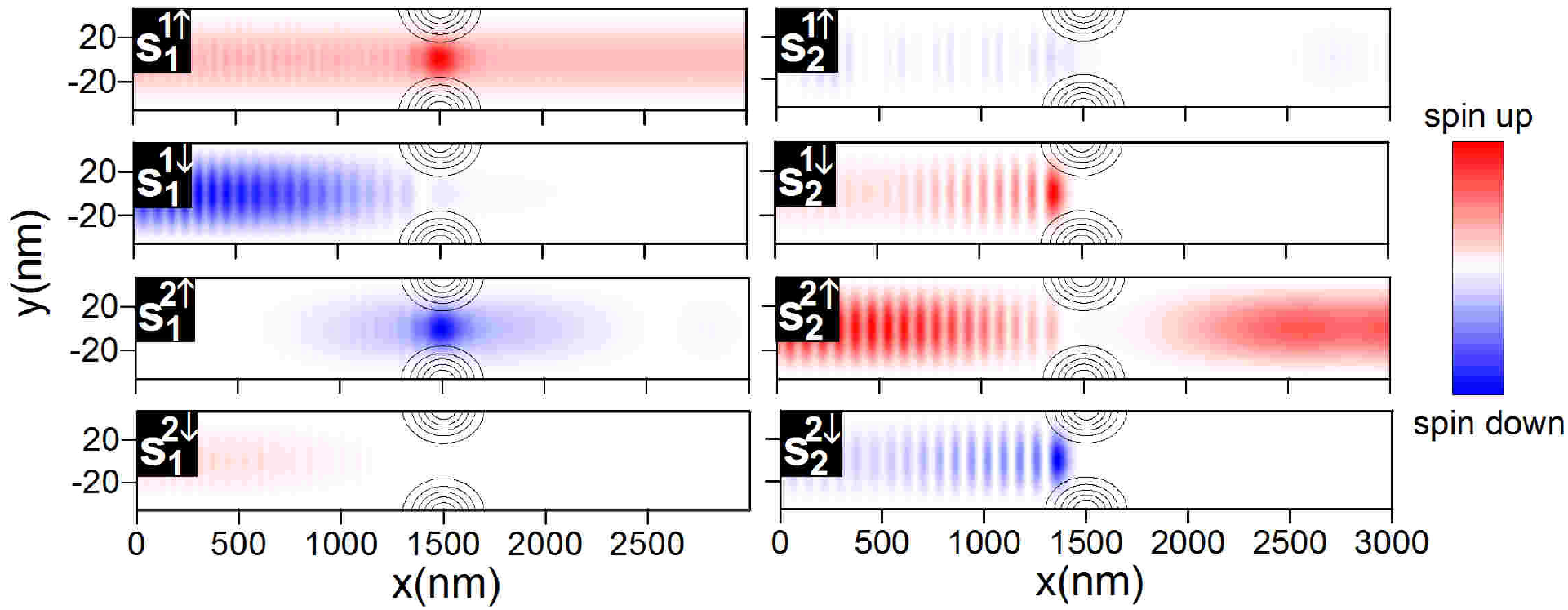}
\caption{  The $z$ component of the partial spin
density distributions $s_{m}^{n\sigma}$, where $m$ is the index of the subband in which the spin
density distribution is presented, while $n\sigma$ denote the index of the subband, including spin,
from which the electrons are injected into the nanowire. The gray contours present the QPC region.
Results for $E_F=3.15$~meV and $V_{QPC}=12$~meV. }
\label{fig6}
\end{center}
\end{figure}

Now, let us explain in detail the spin filtering mechanism, which emerges if we add the QPC.
For this purpose, in Fig.~\ref{fig6} we present the $z$ component of the partial spin density
distribution for $V_{QPC}=12$~meV.
Even a cursory analysis of this figure indicates, that for the chosen Fermi energy, the current
through the QPC is carried mainly by the spin-up electrons transmitted through both the subbands,
in agreement with the spin-dependent transmission probabilities presented in Fig.~\ref{fig3}. For
completeness, we have also calculated the local density of states (LDOS). 
\begin{figure}[ht]
\begin{center}
\includegraphics[scale=0.8, angle=0]{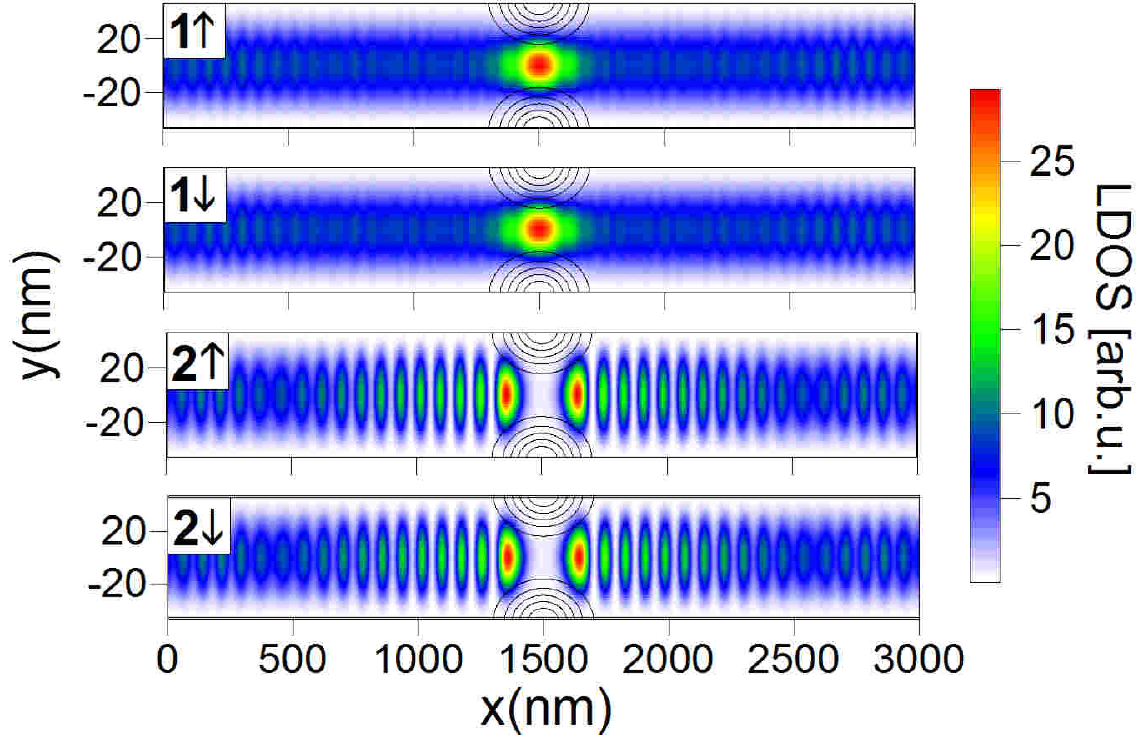}
\caption{ Local density of states (LDOS) calculated for the subbands participating in
the electron transport. The gray contours correspond to the QPC region.}
\label{fig7}
\end{center}
\end{figure}
Figure~\ref{fig7} shows that there are no available states $|2,\sigma \rangle$
($\sigma=\uparrow,\downarrow$) in the QPC region, which means that the electrons reaching QPC  in
the state $|2, \sigma \rangle$, independently of spin, are reflected from it. Now, we consider
separately the transport processes for the
electrons injected into the nanowire from the subsequent subbands (i) $|1,\uparrow \rangle$,
(ii) $|1,\downarrow \rangle$, (iii) $|2,\uparrow \rangle$ and (iv) $|2,\downarrow \rangle$.
Our analysis will be conducted on the basis of the partial spin density distributions
(Figs.~\ref{fig2} and \ref{fig6}), LDOS (Fig.~\ref{fig7}) and the dispersion relations in different
parts of the nanowire presented in Fig.~\ref{fig8}(a).
\begin{figure*}[ht]
\begin{center}
\includegraphics[scale=0.8, angle=0]{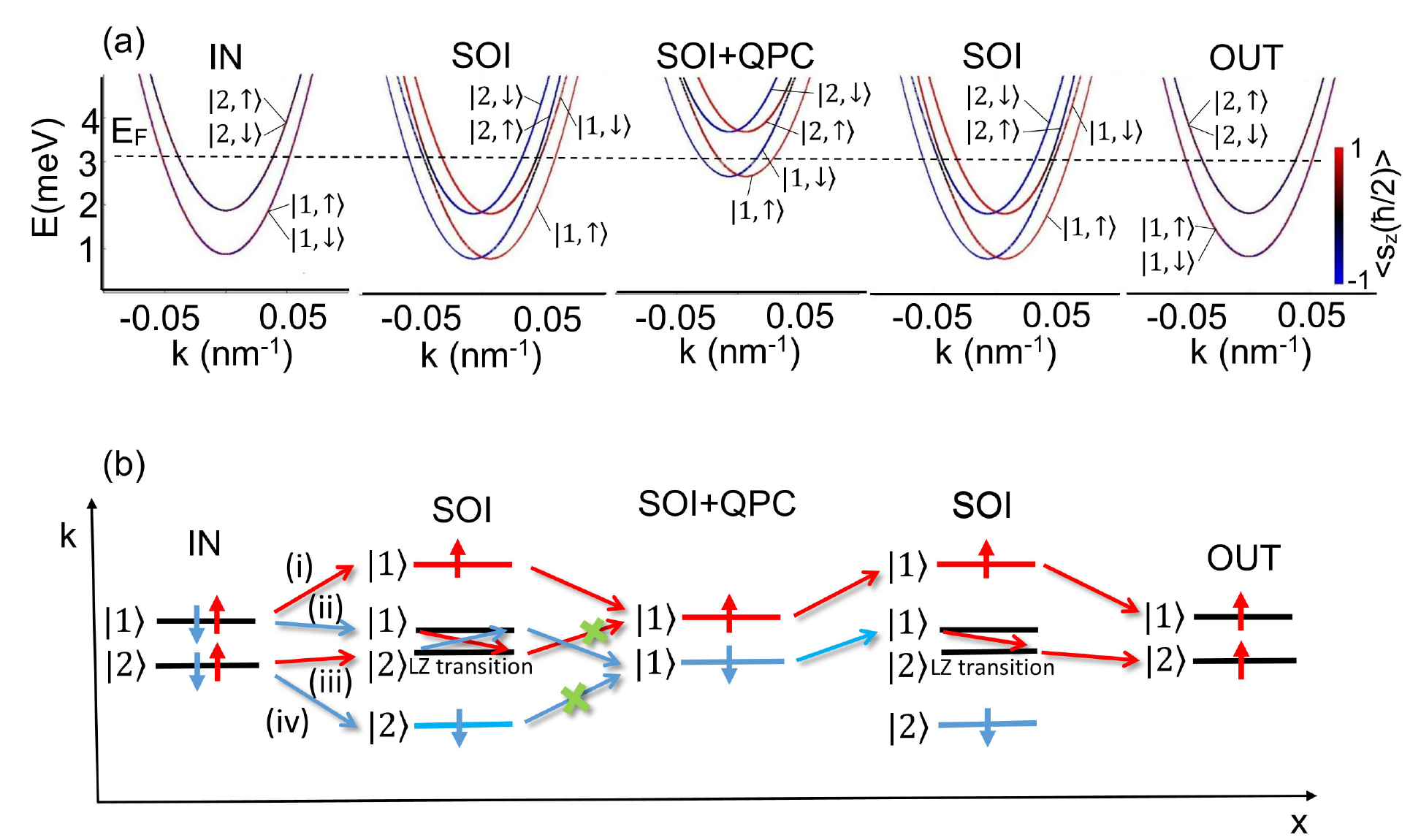}
\caption{  (a) Dispersion relation $E(k)$ in the leads IN and OUT as well as in
the nanowire with the SO interaction and in the QPC region. The dashed horizontal
line denotes the Fermi energy for which the spin density distribution maps are presented in
Fig.~\ref{fig6}. (b) Schematic illustration of the possible transmission processes through the
nanowire.}
\label{fig8}
\end{center}
\end{figure*}
\begin{figure}[ht]
\begin{center}
\includegraphics[scale=0.8, angle=0]{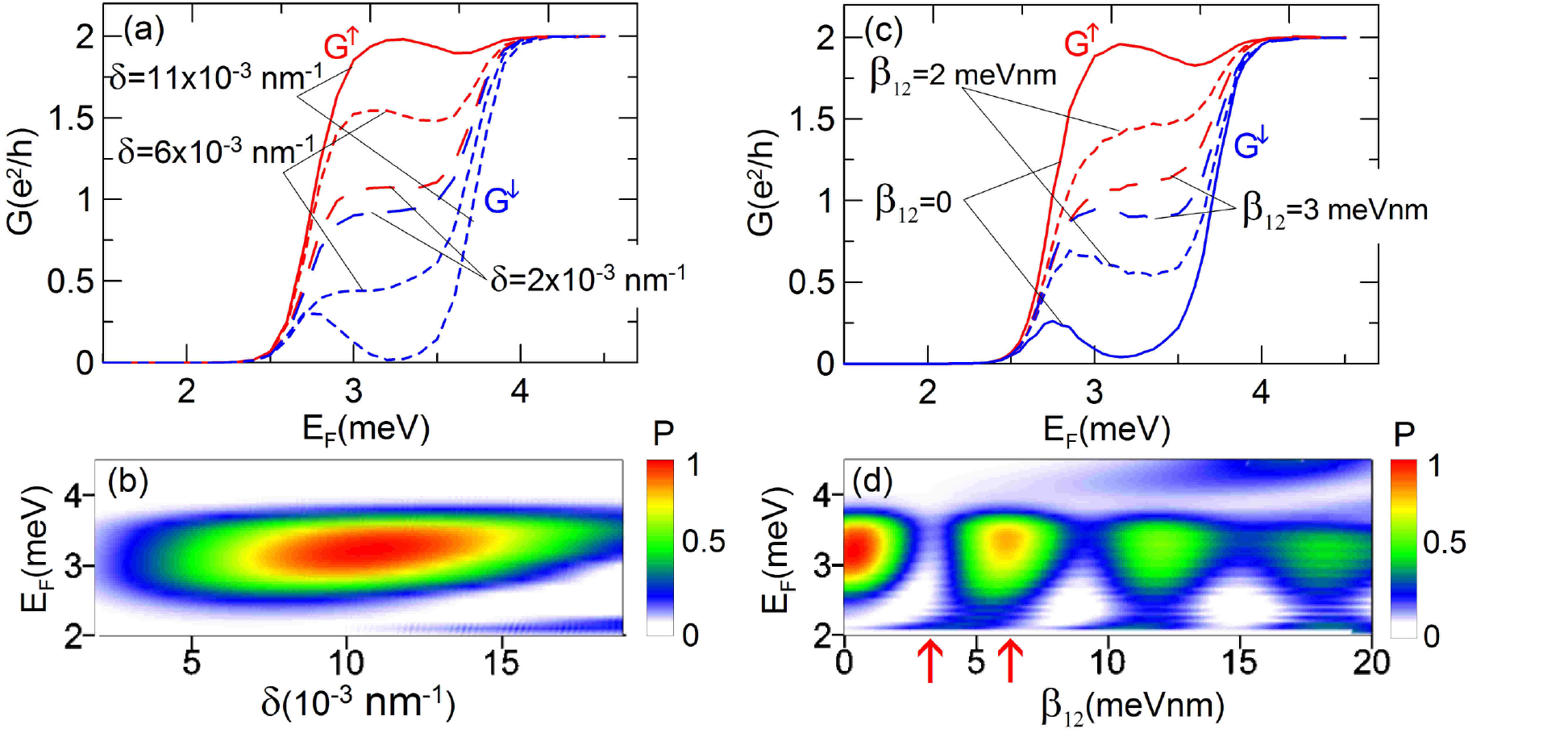}
\caption{
(a) Conductance $G^{\uparrow,\downarrow}$ as a function of Fermi energy $E_F$ for different values
of inter-subband coupling constant $\delta$. Red and blue lines correspond to $G^{\uparrow}$ and
$G^{\downarrow}$, respectively. (b) Spin polarization of current $P$ as a function of Fermi energy
$E_F$ and inter-subband coupling constant $\delta$. (c) Conductance
$G^{\uparrow,\downarrow}$ as a function of Fermi energy $E_F$ for different values
of inter-subband induced SO coupling constant $\beta _{12}$ and $\delta=10^{-2}$~nm$^{-1}$. (d)
Spin polarization of current $P$ as a function of Fermi energy $E_F$ and inter-subband induced SO
coupling constant $\beta _{12}$. The values of $\beta_{12}$ marked by the red arrows are taken to
the further analysis. Results for $V_{QPC}=12$~meV.}
\label{fig9}
\end{center}
\end{figure}

First, we focus on the processes (i) and (iv). As shown in Fig.~\ref{fig4}(b),
which presents the spin dynamics in the nanowire without QPC, the electrons injected in the state
$|1,\uparrow \rangle$ flow through the nanowire conserving their state. Due to the large density of
states $|1,\uparrow \rangle$ in the QPC region (see Fig.~\ref{fig7}), the electrons
injected in this subband are transmitted through the QPC giving raise to the spin-up polarized
current in the output (see Fig.~\ref{fig6}). (iv) Although the electrons in the state $|2,\downarrow
\rangle$ injected into the nanowire without QPC also flow through the nanowire conserving the state
[Fig.~\ref{fig4}(b)], if we introduce the QPC, the electrons are backscattered from the
constriction due to the lack of the states $|2,\downarrow \rangle$ in the QPC region as presented in
Fig.~\ref{fig7}.

The spin dynamics occurring for the electrons injected in the states
$|1,\downarrow \rangle$ and $|2,\uparrow \rangle$ is much more complicated. (ii) Due to the
hybridization of the subbands
$|1,\downarrow \rangle$ and $|2,\uparrow \rangle$ induced by the SO interaction (see avoiding
crossing in Fig.~\ref{fig8}(a)), before reaching the QPC the electrons injected in the state $|1,
\downarrow \rangle$ are transmitted to the state $|2,\uparrow \rangle$. This process is similar
to the previously explained spin dynamics, which occur in the nanowire without QPC
[Fig.~\ref{fig4}(b)]. Therefore, when reaching the QPC the electrons initially injected
in $|1, \downarrow \rangle$ are in the state $|2, \uparrow \rangle$ for which LDOS is zero
in the QPC region (Fig.~\ref{fig7}). These electrons are backscattered from the QPC as presented in
Fig.~\ref{fig6}.
The enhancement of the spin polarization of current is mainly related to the electrons
injected in the state $|2, \uparrow \rangle$ (iii). Although there are no available states for the
subband $|2, \uparrow \rangle$ in the QPC region (Fig.~\ref{fig7}), the electrons in the subband
$|2, \uparrow \rangle$ injected from the contact IN into the nanowire are transmitted to the state
$|1,\downarrow \rangle$ before reaching the QPC (see Fig.~\ref{fig6}). The Landau-Zener
transition is possible due to the lateral SO interaction, which causes the hybridization of the
states $|1,\downarrow \rangle$ and $|2,\uparrow \rangle$. Since LDOS in the QPC region for the state
$|1, \downarrow \rangle$ is large (Fig.~\ref{fig7}), the electrons pass through the QPC and, as
show in Fig.~\ref{fig6}, just after passing through the QPC they are transmitted to the state
$|2, \uparrow \rangle$ leaving the nanowire in this subband. This process, together with the
direct transition through the QPC in the state $|1, \uparrow \rangle$ result in the nearly full
spin-up polarization of the current for the chosen Fermi energy. The transport precesses (i)-(iv)
are schematically illustrated in Fig.~\ref{fig8}(b).
\begin{figure}[ht]
\begin{center}
\includegraphics[scale=0.9, angle=0]{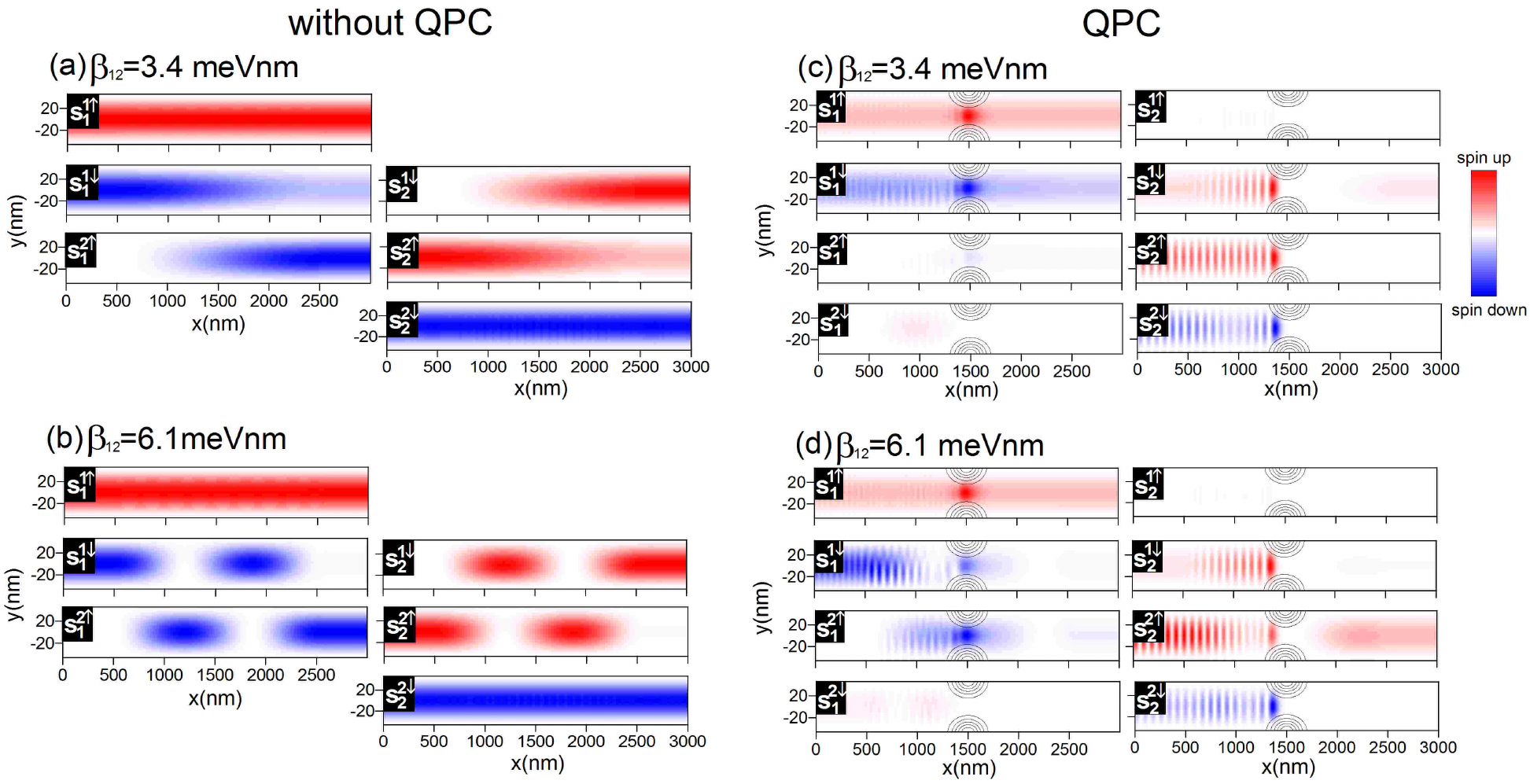}
\caption{ The $z$ component of the partial spin
density distributions $s_{m}^{n\sigma}$ calculated for the nanowire without QPC ($V_{QPC}=0$), for
(a)~$\beta _{12}=3.4$~meVnm and (b)~$\beta_{12}=6.1$~meVnm and for the nanowire with QPC
($V_{QPC}=12$~meV), for (c)~$\beta _{12}=3.4$~meVnm and (d)~$\beta_{12}=6.1$~meVnm. The gray
contours in figures (c) and (d) correspond to the QPC region.
Results for $E_F=3.15$~meV. }
\label{fig10}
\end{center}
\end{figure}

Based on the above analysis one can determine three important factors
necessary to achieve the spin polarization of current in the bilayer nanowires: (a) the lateral
Rashba SO interaction, (b) the inter-subband coupling and (c) the constriction, which in our case
has the form of the QPC. Among these factors the inter-subband coupling
resulting from the lateral Rashba SO interaction plays the crucial role. In order to show how the
value of the parameter $\delta$ affects the spin polarization of current, in Fig.~\ref{fig9}(a) we
present $G^{\uparrow \downarrow}(E_F)$ for
different value of $\delta$. The map of the spin polarization $P$ as a function of the Fermi
energy $E_F$  and the inter-subband coupling constant $\delta$ is presented in Fig.~\ref{fig9}(b).
These results show that there is a value of $\delta$, for which the spin polarization is
the largest and reaches $P=1$.

All the results presented so far have been obtained for the inter-subband SO coupling
constant $\beta_{12}=0$. However, the recent theoretical studies\cite{Calsaverini2008}, in which
the self-consistent calculations of the intra- and inter-subband SO coupling constants were
performed for the double quantum wells, reported that the inter-subband SO coupling constant
exhibits the resonant behavior reaching the values comparable to the ordinary (intra-subband) Rashba
SO constant. For this reason we introduce the inter-subband SO
interaction to our model. Figure~\ref{fig9}(c,d) presents the conductance 
$G^{\uparrow \downarrow}(E_F)$ for different values of the inter-subband SO coupling constants
$\beta _{12}$ assuming $\delta=10^{-2}$~nm$^{-1}$.
The dependence  of the spin polarization $P$ on $E_F$ and $\beta _{12}$
presented in Fig.~\ref{fig9}(d) reveals the damped oscillations of $P$ as a function of $\beta
_{12}$ with the period approximately equal to $6$~meVnm. The inter-subband SO interaction
causes that the electrons flowing through the nanostructure are periodically transfered  between
the hybridized states
$|1,\downarrow \rangle$ and $|2,\uparrow \rangle$. Each of these transitions takes place at a
specific distance, that depends on the coupling constant $\beta _{12}$. Depending of the number of
transitions, which occur when the electrons pass the distance from the lead IN to the QPC region
we obtain the maximal (for odd number of transitions) or minimal (for even number of transitions)
spin polarization. In order to illustrate this property, in Fig.~\ref{fig10} we present the $z$
component of the partial spin density distributions $s_{m}^{n\sigma}$ for two chosen inter-subband
induced SO coupling constants: (a) $\beta _{12}=3.4$~meVnm corresponding to the minimum and (b)
$\beta _{12}=6.1$~meVnm corresponding to the maximum of $P$.  The chosen values of $\beta_{12}$ are
marked by the red arrows in Fig.~\ref{fig9}(d). 
Since our goal is to analyze the
inter-subband spin dynamics under the influence of the inter-subband SO interaction and its impact
on the spin filtering effect, results in Fig.~\ref{fig10} are presented for the two cases, for the
nanowire without QPC ($V_{QPC}=0$) (a,b) and with QPC
($V_{QPC}=12$~meV) (c,d).

Comparing the spin dynamics for the nanowire without QPC presented in Fig.~\ref{fig4}
and Figs.~\ref{fig10}(a,b) one can conclude that the only difference, which appears when we include
the inter-subband induced SO interaction, is the distance that the electron injected in the state 
$|1,\downarrow \rangle$ ($|2,\uparrow \rangle$) needs to make a transition to the state 
$|2,\uparrow \rangle$ ($|1,\downarrow \rangle$). Moreover, independently of the inter-subband SO
coupling $\beta _{12}$, the electrons injected in the states $|1,\uparrow \rangle$ and
$|2,\downarrow \rangle$ remain in their states flowing through the nanowire.
As explained, the spin filtering mechanism presented in the paper requires that the electrons
injected from the contact IN in state $|1,\downarrow \rangle$ ($|2,\uparrow \rangle$) make
transitions to the other state and reach the QPC region in state $|2,\uparrow \rangle$
($|1,\downarrow
\rangle$). This necessary condition causes that the electrons injected in the state $|1,\downarrow
\rangle$ are backscattered from the QPC while the electrons in the state $|2,\uparrow \rangle$
pass through the QPC giving raise to the high spin polarization of the current. As shown in
Fig.~\ref{fig10}(a), for $\beta _{12}=3.4$~meVnm corresponding to the spin polarization minima, the
electrons in the state $|1,\downarrow \rangle$ ($|2,\uparrow \rangle$) are transmitted 
to the state $|2,\uparrow \rangle$ ($|1,\downarrow \rangle$) after traveling the distance $L$. Since
the QPC is located at $x=L/2$, the electrons injected from the lead IN reach the QPC in the state
that is mostly built from the initial state. This means that the spin
filtering mechanism described in the paper will not occur. Instead, as presented in
Fig.~\ref{fig10}(c) the electrons in the states $|1,\uparrow \rangle$ and $|1,\downarrow
\rangle$ simply pass through the QPC while the electrons in the states $|2,\uparrow \rangle$ and
$|2,\downarrow \rangle$ are backscattered due to zero LDOS in the QPC region (see
Fig.~\ref{fig7}). The electron transport in the nanowire through the subbands with
opposite spin leads to the spin polarization nearly equal to zero, i.e. the unpolarized current.
The further increase of the inter-subband induced SO interaction constant causes that the distance
over which the electron are transmitted between the states $|1,\downarrow \rangle$ and $|2,\uparrow
\rangle$ becomes shorter. For $\beta_{12}=6.1$~meVnm [Fig.~\ref{fig10}(b) and (d)] the necessary
condition, for which the spin filtering effect occurs, namely the inter-subband transition before
reaching QPC, is again satisfied giving raise to the high spin polarization of the current.
We can summarize that depending on the number of the inter-subband transitions, which
occur before the electron reaches the QPC acting as the channel selector, the spin polarization
oscillates between the large and small values as presented in Fig.~\ref{fig9}(d).

\begin{figure}[ht]
\begin{center}
\includegraphics[scale=0.9, angle=0]{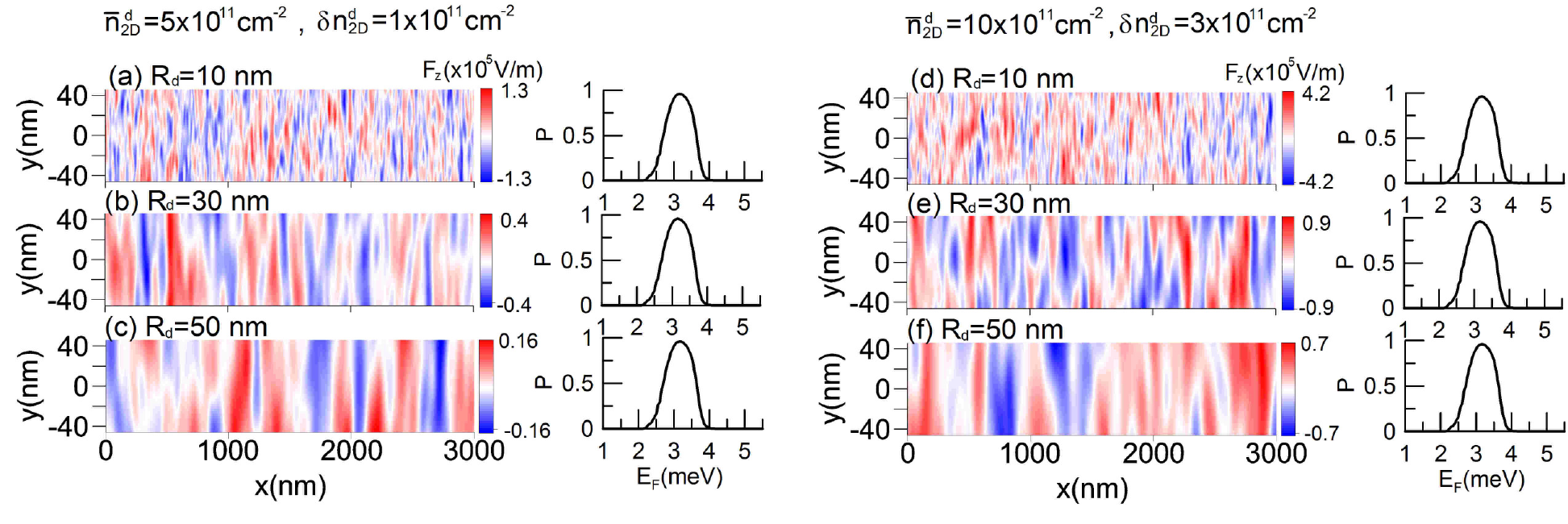}
\caption{Spatial distributions of the electric field $F_z(x,y)$ calculated for the different $\bar{n}^d_{2D}$ and $\delta n^d_{2D}$ as well as
different distances $R_d$ of the dopant layers from the main quantum well. Right panels present the spin polarization of the current $P(E_F)$
calculated with the inclusion of the random component of the spin-orbit coupling generated by the electric field distribution presented on the
left panels.}
\label{fig11}
\end{center}
\end{figure}

Our results in general provide a new mechanism to implement spin-polarized electron sources in the
bilayer nanowires. Therefore, it is important to discuss this proposal from the point of view of the possible perturbations, 
which can appear in the experimental realization of the proposed spin filter.
Our model assumes three necessary conditions, which have to be satisfied in order to achieve the high spin
polarization of the current: (a) the lateral Rashba SO interaction,
(b) the inter-subband coupling and (c) the constriction, which in the present paper has the form of the QPC.
The lateral Rashba SO interaction is the most important requirement because it generates the coupling between the 
subbands, which in the consider nanostructures, is crucial for the spin polarization. However, in the wells made of 
materials with the zincblende crystallographic structure (GaAs, InAs, etc.), the SO coupling also originates from the bulk 
Dresselhaus term. Although in the GaAs-based structures both the Rashba and Dresselhaus terms are often of the same order 
of magnitudes, in the InGaAs-based wells growing in the $[001]$ direction, the Rashba term dominates. 
Since the linear Dresselhaus parameter $\gamma _D \sim 1 / d_{QW}^2$, this domination is strengthened for 
the wide quantum wells needed for the experimental realization of the bilayer nanowires, e.g 
for the gated Al$_{0.48}$In$_{0.52}$As/Ga$_{0.47}$In$_{0.53}$As double quantum well structure, the Dresselhaus SO parameter is 
a four orders of magnitude smaller than the Rashba coupling constant - see the Supplementary material. We have checked by performing direct
calculations that the inclusion of such a small term into our model does not affect the spin filtering presented in the paper. 
Although the neglect of the Dresselhaus term in our model is fully justified, the assumption of  constant, spatially independent  
Rashba parameter requires a detailed discussion. In the considerations presented so far, we assume that the Rashba SO parameter is constant and
spatially independent.
However, in the realistic nanostructure any local imperfection leads to the local change in the SO coupling. Since, the carriers in the quantum well
come from donors, the electric field of ionized donors generates the random, spatially dependent electrostatic potential in the quantum well. 
This causes that the SO interaction has a random component, which for some structures can be large. Such strong fluctuations of
the Rashba coupling constant have been recently measured by the scanning tunnelling microscopy in the structure with 2DEG fabricated by adsorbing
Cs  on the p-type InSb(110)\cite{Bindel2016}. In that case\cite{Bindel2016}, the strong spatial fluctuations of the Rashba parameter between
$0.4$~eV$\AA$ and $1.6$~eV$\AA$ result from the particular, random locations of the dopant ions, which are very close to the inversion layer
where the electron density is located. In the
Al$_{0.48}$In$_{0.52}$As/Ga$_{0.47}$In$_{0.53}$As double quantum well structure suggested to the experimental realization of
the proposed spin filter, the donor-doped layers are located symmetrically on both sides of the well in the distance $R_d$ much larger than the their
width $w_d$ (usually $R_d/w_d \approx 5-10$).The random distribution of the electric field and the Rashba spin-orbit parameter in the main quantum
well originating from the inhomogeneity of the dopant concentration in the donor-doped layers has been discussed in detail in Ref.~\cite{Glazov2010}.
Following this model\cite{Glazov2010} we have studied the role of the random spin-orbit coupling component generated by the donors
on the spin filter effect presented in the paper. For this purpose, we have calculated the $z$-component 
of the electric field of the dopant ions with the local concentration $n^d_{2D}(x',y')$
\begin{equation}
 F_z(x,y)=-\frac{|e|}{4 \pi \varepsilon _0 \varepsilon} \int n^d_{2D}(x',y')\frac{R_d}{[(x-x')^2+(y-y')^2+R_d^2]^{3/2}} dx' dy',
\end{equation}
where $e$ is the electron charge and $\varepsilon$ is the dielectric constant. We assume that the random
distribution of dopants $n^d_{2D}(x',y')$ obeys the Gaussian statistics with the mean value $\bar{n}^d_{2D}$ and the standard deviation $\delta
n^d_{2D}$. In Fig.~\ref{fig11} we present the spatial distributions of the electric field $F_z(x,y)$ calculated for the different mean values of donor
concentration $\bar{n}^d_{2D}$ and deviations $\delta n^d_{2D}$ as well as different distances $R_d$ of the dopant layers from the main
quantum well. We assume $\varepsilon=14$ corresponding to InAs.
The spatially dependent electric field $F_z(x,y)$ leads to the random distribution of the intra-subband SO coupling 
$\beta_{11}$ and $\beta_{22}$ [see hamiltonian (5) in the Supplementary material] which in the presented model were assumed to be zero due to the $z
\leftrightarrow -z$ symmetry. In order to show how this effect affects the spin-filtering in the considered nanostructures, we include the
random component of the spin-orbit coupling into our model and calculate the spin polarization of the current $P$ as a function of the Fermi energy
$E_F$. The right panels in Fig.~\ref{fig11} present the spin polarization of the current $P(E_F)$ calculated with the inclusion of the spatial
distribution of the intra-subband spin-orbit parameters, generated by the electric field distribution presented on the left panels.
In the calculations we assume $\beta_{11}(x,y)=\beta_{3D}|e|F_z(x,y)$ (we take on $\beta_{3D}=0.572$~nm$^{-2}$ corresponding to InAs) and
$\beta_{22}(x,y)=-\beta_{11}(x,y)$. The second assumption is based on our
recent results\cite{Wojcik2016_arxiv} which show that the intra-subband spin-orbit coupling constants for the ground and first excited state have
nearly the same value but opposite sign. The rest of the parameters are assumed to be the same as used in the calculations presented the
Fig.~\ref{fig2}. Figure \ref{fig11} shows that for the reasonable values of the dopant concentration and its location with respect to the main quantum
well, the random spatial distribution of the spin-orbit coupling does
not affect the spin filtering presented in the paper. Its negligible contribution results from the fact that the $z$-component of the electric field
of the ionised dopants is several orders of magnitude smaller than the lateral electric field needed to obtain the spin filter effect and used in the
experiments for modulation of the Rashba parameter by the QPCs~\cite{Das2011}. In the above discussion we did not consider the screening of the
$z$-component of the electric field  due to the $z \leftrightarrow -z$ symmetry\cite{Glazov2010}. As shown in Ref.~\cite{Glazov2010}, the screening is
important for the in-plane components of the electric field, however, the estimated 
ratio of the fluctuations of the lateral and transversal components of the electric field~\cite{Glazov2010} in the symmetrically doped quantum well
is $F_{||} / F_{z} \ll 1$. Therefore, the screening strongly reduces the fluctuations of the lateral electric field and (as much smaller
than the transversal one) does not affect the spin filter effect presented in the paper. 

\section*{Summary}
The inter-subband SO interaction attracts the growing interest, because it gives
raise to interesting physical effects, e.g., unusual Zitterbewegung\cite{Bernardes2007}. In
the bilayer nanowires, the strength of this specific SO interaction, arising from the coupling
between states with opposite parity, is comparable to the Rashba intra-subband SO coupling.
It makes the bilayer nanowires  a good candidate for investigating the inter-subband SO interaction
and the effects related with it. 

In the present paper, we have proposed the spin filtering mechanism based on the
inter-subband SO interaction in the bilayer nanowire with QPC. For this purpose we have studied the
electron transport through the nanowire within the two-subband model including the inter-subband SO
interaction induced by the lateral electric field. We have found that for the non-zero inter-subband
coupling, in the presence of the lateral Rashba SO interaction, the current flowing through
the QPC is almost fully spin polarized. In order to explain the spin filtering effect, first we
have considered the bilayer nanowire without QPC. By the use of the partial spin density
distributions calculated for each of the subband participating in the transport, we have shown that 
the electrons injected in the state $|1,\downarrow \rangle$ ($|2,\uparrow \rangle$) are transmitted
to the state $|2, \uparrow \rangle$ ($|1,\downarrow \rangle$) in the middle of the nanowire and
again are back in their original state before leaving the nanowire. On the other hand, the
electrons in the states $|1,\uparrow \rangle$  and $|2,\downarrow \rangle$ flow through the nanowire
conserving their state. The observed Landau-Zener transitions are caused by the
hybridization of states $|1,\downarrow \rangle$ and $|2,\uparrow \rangle$ induced by the
lateral Rashba SO interaction, which mixes the orbital and spin degrees of freedom.
The proposed spin filtering mechanism emerges after adding the QPC, which blocks the
electron traveling in the states $|2, \sigma \rangle$ - as shown in Fig.~\ref{fig7}, LDOS for these
states in the QPC region is zero. Therefore, the introduction of the QPC causes that only the
electrons injected into the states $|1,\uparrow \rangle$ and $|2,\uparrow \rangle$ are transmitted
through the QPC giving raise to the high spin polarization of the current - the electrons in the
state $|1,\uparrow \rangle$ pass through the QPC remaining in their state while the electrons
injected in  $|2,\uparrow \rangle$ are transmitted to the state $|1,\downarrow \rangle$ before
reaching QPC, pass through the QPC in the state $|1,\downarrow \rangle$ and just behind the QPC
 are again transmitted to  $|2,\uparrow \rangle$ leaving the nanowire in this state. 
Summing up, the proposed spin filtering effect can be explained as the combined effect of the
Landau-Zener inter-subband transitions caused by the
hybridization of states with opposite spin and the
confinement in the QPC region. We
have determined the three important factors necessary to achieve the high spin polarization of the
current in the bilayer nanowire: (a) the lateral Rashba SO interaction,
(b) the inter-subband coupling and (c) the constriction which in our case has the form of the QPC.

Our results in general provide a new mechanism to implement spin-polarized electron sources in the
realistic bilayer nanowires with QPC, which can be realized
experimentally in the double quantum well structure or the wide quantum well. This is especially 
interesting in the context of the current research on the spin filtering in QPC with single
occupancy\cite{Debray2009,Das2011,Das2012,Bhandari2012,Bhandari2013,Wan2009,Bhandari2013,
Kohda2012}. In those systems, the lateral Rashba SO interaction causes a small spin
imbalance, which then is gained by the electron-electron interaction. Our proposal, based on the
inter-subband SO interaction in the bilayer nanowires, gives the nearly full spin
polarization even without inclusion of the electron-electron interaction. This allows us to expect
that the proposed spin filtering effect is more efficient and in the near future 
can lead to the fabrication of the efficient spin filter.


\section*{Acknowledgements}
This work was supported by the funds of Ministry of Science and Higher Education for 2016 and by
PL-Grid Infrastructure.

\section*{Author contributions statement}
P.W. conceived the idea, performed the computer simulations, analyzed and interpreted the data and wrote the paper, J.A. corrected the paper.

\section*{Additional Information}
\textbf{Supplementary information} accompanies this paper. \\
\textbf{Competing financial interests:} The authors declare no competing financial interests. \\

\end{document}